\documentstyle[12pt,epsfig]{article}
\newlength{\egwidth}
\newenvironment{eg}%
{\begin{list}{}{\setlength{\leftmargin}{0.05\textwidth}%
\setlength{\rightmargin}{\leftmargin}}\item[]\normalsize}%
{\end{list}}
\newenvironment{egbox}%
{\begin{minipage}[t]{\egwidth}}%
{\end{minipage}}
\newcommand{\egstart}{\begin{eg}\begin{egbox}}

\newcommand{\egend}{\end{egbox}\end{eg}}

\newcommand{\bea}{\begin{eqnarray}}
\newcommand{\eea}{\end{eqnarray}}
\newcommand{\be}{\begin{equation}}
\newcommand{\ee}{\end{equation}}

\newcommand{\nn}{\nonumber}
\newcommand{\rf}[1]{(\ref{#1})}

\begin{document}

\begin{center}
{\large\bf Existence of the $\sigma$-Meson below 1 GeV and the $f_0(1500)$
Glueball} \\
\bigskip

\underline{\bf Yu.S.~Surovtsev}$^{a,}$\footnote{E-mail address:
surovcev@thsun1.jinr.ru},
{\bf D.~Krupa}$^b$, {\bf M.~Nagy}$^b$\\
$^a${\sl Bogoliubov Laboratory of Theoretical Physics, Joint Institute for
Nuclear Research, Dubna 141 980, Moscow Region, Russia},\\
$^b${\sl Institute of Physics, Slov.Acad.Sci., D\'ubravsk\'a cesta 9,
842 28 Bratislava, Slovakia}
\end{center}
\begin{abstract}
On the basis of a simultaneous description of the isoscalar $s$-wave channel
of the $\pi\pi$ scattering (from the threshold up to 1.9 GeV) and of the
$\pi\pi\to K\overline{K}$ process (from the threshold to $\sim$ 1.4 GeV) in
the model-independent approach, a confirmation of the $\sigma$-meson at
$\sim$ 665 MeV and an indication for the glueball nature of the $f_0(1500)$
state are obtained. It is shown that the large $\pi\pi$-background, usually
obtained, combines, in reality, the influence of the left-hand branch-point
and the contribution of a very wide resonance at $\sim$ 665 MeV. The coupling
constants of the observed states with the $\pi\pi$ and $K\overline{K}$ systems
and lengths of the $\pi\pi$ and $K\overline{K}$ scattering are obtained.

PACS: 14.40.Cs,11.80.Gw,12.39.Pn,13.75.Lb

{\it Keywords:} Multichannel states; Riemann surface; Pole clusters
\end{abstract}

\newpage

{\bf 1.} A problem of scalar mesons is most troublesome and long-lived in the light
meson spectroscopy. Among difficulties in their understanding
there is the one related to a strong model-dependence of information on
multichannel states obtained in analyses based on the specific dynamic models
or using an insufficiently-flexible representation of states ({\it e.g.}, the
standard Breit -- Wigner form). Earlier, we have shown \cite{KMS-nc96} that an
inadequate description of multichannel states gives not only their distorted
parameters when analyzing data but also can cause the fictitious states when
one neglects important (even energetic-closed) channels. In this paper we are
going, conversely, to demostrate that the large background ({\it e.g.}, that
happens in analyzing $\pi\pi$ scattering), can hide low-lying states, even
such important for theory as a $\sigma$-meson \cite{PDG-98}. The latter is
required by most of the models (like the linear $\sigma$-models or the
Nambu -- Jona-Lasinio models \cite{NJL}-\cite{Scadron}) for spontaneous
breaking of chiral symmetry.
Recent new analyses of the old and new experimental data found a possible
candidate for that state \cite{Zou}-\cite{Kaminski}. However, these analyses
use either the Breit -- Wigner form (even if modified) or specific forms of
interactions; therefore, there one cannot talk about a model independence of
results. Besides, in these analyses, a large $\pi\pi$-background is obtained.

A model-independent information on multichannel states can be obtained on the
basis of the first principles (analyticity, unitarity) immediately applied to
analyzing experimental data. The way of realization is a consistent allowance
for the nearest singularities on all sheets of the Riemann surface of the
$S$-matrix. Earlier, we have proposed this method for 2- and 3-channel
resonances and developed the concept of standard clusters
(poles on the Riemann surface) as a qualitative characteristic of a state
and a sufficient condition of its existence \cite{KMS-nc96,KMS-yf86}.
The cluster kind is related to the state nature. At all events, we can,
in a model-independent manner, discriminate between bound states of particles
and the ones of quarks and gluons \cite{KMS-nc96,MP-92}, qualitatively
predetermine the relative strength of coupling of a state with the considered
channels, and obtain an indication on its gluonium nature.

{\bf 2.} Here we restrict ourselves to a 2-channel simultaneous consideration
of the coupled processes $\pi\pi\to \pi\pi,K\overline{K}$. Therefore, we have
the 2-channel $S$-matrix determined on the 4-sheeted Riemann surface.
The matrix elements $S_{\alpha\beta}$, where $\alpha,\beta=1(\pi\pi),
2(K\overline{K})$, have the right-hand cuts along the real axis of the
$s$-plane, starting at $4m_\pi^2$ and $4m_K^2$, and the left-hand cuts,
beginning at $s=0$ for $S_{11}$ and at $4(m_K^2-m_\pi^2)$ for $S_{22}$ and
$S_{12}$. We number the Riemann-surface sheets according to the signs of
analytic continuations of the channel momenta
~$k_1=(s/4-m_\pi^2)^{1/2}, ~k_2=(s/4-m_K^2)^{1/2}$~ as follows: ~
signs $({\mbox{Im}}k_1,{\mbox{Im}}k_2)=++,-+,--,+-$ correspond to the sheets
I, II, III, IV.

To elucidate the resonance representation on the Riemann surface, we express
analytic continuations of the matrix elements to the unphysical sheets
$S_{\alpha\beta}^L$ ($L=II,III,IV$) in terms of them on the physical sheet
$S_{\alpha\beta}^I$. The latter have, except for the real axis, only zeros
corresponding to resonances. Using the reality property of the analytic
functions and the 2-channel unitarity, one can obtain
\bea \label{S_L}
&&S_{11}^{II}=\frac{1}{S_{11}^I},\qquad ~~~~S_{11}^{III}=\frac{S_{22}^I}{\det S^I},
\qquad S_{11}^{IV}=\frac{\det S^I}{S_{22}^I},\nn\\
&&S_{22}^{II}=\frac{\det S^I}{S_{11}^I},\qquad S_{22}^{III}=\frac{S_{11}^I}
{\det S^I},\qquad S_{22}^{IV}=\frac{1}{S_{22}^I},\\
&&S_{12}^{II}=\frac{iS_{12}^I}{S_{11}^I},\qquad ~~~S_{12}^{III}=\frac{-S_{12}^I}
{\det S^I},\qquad S_{12}^{IV}=\frac{iS_{12}^I}{S_{22}^I},\nn
\eea
Here $\det S^I=S_{11}^I S_{22}^I-(S_{12}^I)^2$.
These formulae immediately give the resonance representation by poles and
zeros on the 4-sheeted Riemann surface. One must discriminate between three
types of 2-channel resonances described by a pair of conjugate zeros on sheet
I: ({\bf a}) in $S_{11}$, ({\bf b}) in $S_{22}$, ({\bf c}) in each of $S_{11}$
and $S_{22}$.
As is seen from \rf{S_L}, to the resonances of types ({\bf a}) and ({\bf b})
one has to make correspond a pair of complex conjugate poles on sheet III,
shifted relative to a pair of poles on sheet II and IV, respectively.
To the states of type ({\bf c}) one must make correspond two pairs of
conjugate poles on sheet III.
A resonance of every type is represented by a pair of complex-conjugate
clusters (of poles and zeros on the Riemann surface) of size typical of
strong interactions.
The cluster kind is related to the state nature. The resonance, coupled
strongly with the $\pi\pi$ channel, is described by the cluster of type
({\bf a}); the resonance, coupled strongly with the $K\overline{K}$ and weakly
with $\pi\pi$ channel (say, an $s\overline{s}$ state), by the
cluster of type ({\bf b}); the flavour singlet ({\it e.g.} glueball) must be
represented by the cluster of type ({\bf c}) as a necessary condition.

For the simultaneous analysis of experimental data on the coupled processes
it is convenient to use the Le Couteur-Newton relations \cite{LC,Newton}
expressing the $S$-matrix elements of all coupled processes in terms of the
Jost matrix determinant $d(k_1,k_2)$, the real analytic function with the only
square-root branch-points at $k_i=0$.
Earlier, this was done by us in the 2-channel consideration \cite{KMS-yf86}
with the uniformizing variable $z=(k_1+k_2)/\sqrt{m_K^2-m_\pi^2}$
which was proposed in Ref.\cite{Kato} and maps the 4-sheeted Riemann surface
with two unitary cuts, starting at $4m_\pi^2$ and $4m_K^2$, onto the plane.
Note that other authors have been also applied the parametrizations with using
the Jost functions at analyzing the $s$-wave $\pi\pi$ scattering in the
one-channel approach \cite{Bohacik} and in the 2-channel one \cite{MP-92}. In
latter work, the uniformizing variable $k_2$ has been used, therefore, their
approach cannot be emploied near by the $\pi\pi$ threshold.

When analyzing the processes $\pi\pi\to \pi\pi,K\overline{K}$ by the above
methods in the 2-channel approach, two states ($f_0 (975)$ and $f_0 (1500)$)
were found to be sufficient ($\chi^2/\mbox{ndf}\approx1.00$).
However, there the large $\pi\pi$-background has been obtained. A character of
the representation of the background (the pole of 2nd order on the
real axis on sheet II and the corresponding zero on sheet I) suggests
that a wide light state is possible to be hidden in the background. To check
this, one must work out the background in some detail.

Now we will take, in the uniformizing variable, into account also the
left-hand branch-point at $s=0$. We use the uniformizing variable
\be \label{v}
v=\frac{m_K\sqrt{s-4m_\pi^2}+m_\pi\sqrt{s-4m_K^2}}{\sqrt{s(m_K^2-m_\pi^2)}},
\ee
which maps the 4-sheeted Riemann surface, having (in addition to two
above-indicated unitary cuts) also the left-hand cut starting at
$s=0$, onto the $v$-plane. In Fig.1, the plane of the uniformizing variable
$v$ for the $\pi\pi$-scattering amplitude is depicted. The Roman numerals
(I, \ldots, IV) denote the images of the corresponding sheets; the thick line
represents the physical region; the points i, 1 and
$b=\sqrt{(m_K+m_\pi)/(m_K-m_\pi)}$ correspond to the $\pi\pi, K\overline{K}$
thresholds and $s=\infty$, respectively; the shaded intervals
$(-\infty,-b],[-b^{-1},b^{-1}],[b,\infty)$ are the images of the
corresponding edges of the left-hand cut.
The depicted positions of poles ($*$) and of zeros ($\circ$) give the
representation of the type ({\bf a}) resonance in $S_{11}$.

On $v$-plane the Le Couteur-Newton relations are \cite{KMS-yf86,Kato}
\be \label{v:C-Newton}
S_{11}=\frac{d(-v^{-1})}{d(v)},\quad S_{22}=\frac{d(v^{-1})}{d(v)},
\quad S_{11}S_{22}-S_{12}^2=\frac{d(-v)}{d(v)}.
\ee
The condition of the real analyticity implies $d(-v^*)=d^* (v)$
for all $v$, and the unitarity needs the following relations to hold true for
the physical $v$-values:
$|d(-v^{-1})|\leq |d(v)|,\quad |d(v^{-1})|\leq |d(v)|,\quad |d(-v)|=|d(v)|.$

The $d$-function that on the $v$-plane already does not possess branch-points
is taken as
~$d=d_B d_{res}$,
where ~$d_B=B_{\pi}B_K$; $B_{\pi}$ contains the possible remaining
$\pi\pi$-background contribution, related to exchanges in crossing channels;
$B_K$ is that part of the $K\overline{K}$ background which does not contribute
to the $\pi\pi$-scattering amplitude. The function $d_{res}(v)$ represents the
contribution of resonances, described by one of three types of the pole-zero
clusters, {\it i.e.}, except for the point $v=0$, it consists of zeros of
clusters:
\be \label{d_res}
d_{res} = v^{-M}\prod_{n=1}^{M} (1-v_n^* v)(1+v_n v),
\ee
where $M$ is the number of pairs of the conjugate zeros.

{\bf 3.} Here we analyze simultaneously the available experimental data on
the $\pi\pi$-scattering \cite{Hyams} and the process $\pi\pi\to K\overline{K}$
\cite{Wickl} in the channel with $I^GJ^{PC}=0^+0^{++}$.

To obtain the satisfactory description of the $\pi\pi$ scattering
from the threshold to 1.89 GeV, we have taken $B_\pi=1$, and three states
turned out to be sufficient: the two ones of the type ({\bf a}) ($f_0 (665)$
and $f_0 (980)$) and $f_0 (1500)$ of the type ({\bf c}). The following zero
positions on the $v$-plane, corresponding to these resonances, have been
established:
\bea
{\rm for} ~~f_0 (665):
~~&&v_1=1.36964+0.208632i,\qquad v_2 =0.921962-0.25348i,\nn\\
{\rm for} ~~f_0 (980):~
&&v_3=1.04834+0.0478652i,\qquad ~v_4 =0.858452-0.0925771i,\nn\\
{\rm for} ~~f_0 (1500):
&&v_5=1.2587+0.0398893i,\qquad ~~~v_6 =1.2323-0.0323298i,\nn\\
&&v_7=0.809818-0.019354i,\qquad v_8 =0.793914-0.0266319i.\nn
\eea
Here for the $\pi\pi$ phase shift $\delta_1$ and the elasticity parameter
$\eta$, 113 and 50 experimental points \cite{Hyams}, respectively, are used;
when rejecting the points at 0.61, 0.65, and 0.73 GeV for $\delta_1$
and at 0.99, 1.65, and 1.85 GeV for $\eta$, which give an anomalously
large contribution to $\chi^2$, we obtain for $\chi^2/\mbox{ndf}$ the values
2.7 and 0.72, respectively; the total $\chi^2/\mbox{ndf}$ in the case of the
$\pi\pi$ scattering is 1.96.

With the presented picture, the satisfactory description for the modulus of
the $\pi\pi\to K\overline{K}$ matrix element $|S_{12}|$ is given from the
threshold to $\sim$ 1.4 GeV (Fig.2). Here 35 experimental points \cite{Wickl}
are used; $\chi^2/\mbox{ndf}\approx 1.11$ when eliminating the points at
1.002, 1.265, and 1.287 GeV (with especially large contribution to
$\chi^2$). However, for the phase shift $\delta_{12}(s)$, slightly excessive
curve is obtained. Therefore, keeping the {\it parameterless} description of
the $\pi\pi$ background, one must take into account the part of the
$K\overline{K}$ background that does not contribute to the
$\pi\pi$-scattering amplitude. Note that on the $v$-plane, $S_{11}$ has
no cuts; however, the amplitudes of the processes
$K\overline{K}\to\pi\pi,K\overline{K}$ do have the cuts which arise from
the left-hand cut on the $s$-plane, starting at $s=4(m_K^2-m_\pi^2)$.  This
left-hand cut will be neglected in the Riemann-surface structure, and the
contribution on the cut will be taken into account in the $K\overline{K}$
background as a pole on the real $s$-axis on the physical sheet in the
sub-$K\overline{K}$-threshold region; on the $v$-plane, this pole gives two
poles on the unit circle in the upper half-plane, symmetric to each other with
respect to the imaginary axis, and two zeros, symmetric to the poles with
respect to the real axis, {\i.e.}, one additional parameter is introduced
(a position $p$ of the pole on the unit circle). Therefore, for $B_K$ we take
the form
\be \label{B_K}
B_K=v^{-4}(1-pv)^4(1+p^*v)^4.
\ee
Fourth power in \rf{B_K} is stipulated by the following. First, a pole on the
real $s$-axis on sheet I in $S_{22}$ is accompanied by a pole on sheet II at
the same $s$-value (as it is seen from eqs. \rf{S_L}); on the $v$-plane, this
implies the pole of second order. Second, for the $s$-channel process
$\pi\pi\to K\overline{K}$, the crossimg $u$- and $t$-channels are the $\pi-K$
and $\overline{\pi}-K$ scattering; this results in the additional doubling of
the multiplicity of the indicated pole on the $v$-plane. The expression
\rf{B_K} does not contribute to $S_{11}$, {\it i.e.} the parameterless
description of the $\pi\pi$ background is kept. A satisfactory description of
the $\delta_{12}(\sqrt{s})$ (Fig.3) is obtained to $\sim$1.52 GeV with the
parameter $p=0.948201+0.31767i$ (that corresponds to the position of the pole
on the $s$-plane at $s=0.434 {\rm GeV}^2$). Here 59 experimental points
\cite{Wickl} are considered; $\chi^2/\mbox{ndf}\approx 3.05$ when eliminating
the points at 1.117, 1.247, and 1.27 GeV. The total $\chi^2/\mbox{ndf}$ for
four analyzed quantities to describe the processes
$\pi\pi\to\pi\pi,K\overline{K}$ is 2.12; the number of adjusted parameters is
17, where they all (except a single relating to the $K\overline{K}$
background) are positions of poles describing resonances.

In Table 1, the obtained poles on the corresponding sheets are cited on the
complex energy plane ($\sqrt{s_r}={\rm E}_r-i\Gamma_r$). Since, for wide
resonances, values of masses and widths are very model-dependent, it is
reasonable to report characteristics of pole clusters which must be rather
stable for various models.

Now we can calculate the constants of the obtained-state couplings with the
$\pi\pi-"1"$ and $K\overline{K}-"2"$ systems through the residues of amplitudes at the
pole on sheet II. Expressing the $T$-matrix via the $S$-matrix as
$S_{ii}=1+2i\rho_i T_{ii},~ S_{12}=2i\sqrt{\rho_1\rho_2} T_{12}$,
where $\rho_i=\sqrt{(s-4m_i^2)/s}$, taking the resonance part of the
amplitude in the form:
$T_{ij}^{res}=\sum_r g_{ir}g_{rj}D_r^{-1}(s)$,
where $D_r(s)$ is an inverse propagator ($D_r(s)\propto s-s_r$), and
denoting the coupling constants with the $\pi\pi$ and $K\overline{K}$
systems through $g_1$ and $g_2$, respectively, we obtain for
$f_0(665)$: $g_1=0.7477\pm 0.095$ GeV and $g_2=0.834\pm 0.1$ GeV, for
$f_0(980)$: $g_1=0.1615 \pm 0.03$ GeV and $g_2=0.438 \pm 0.028$ GeV, for
$f_0(1500)$: $g_1=0.899 \pm 0.093$ GeV.

Let us indicate also scattering lengths calculated in our approach. For the
$K\overline{K}$ scattering, we obtain
$a_0^0(K\overline{K})=-1.188\pm 0.13+(0.648\pm 0.09)i,~m_{\pi^+}^{-1}.$
A presence of the imaginary part in $a_0^0(K\overline{K})$ reflects the fact,
that already at the threshold of the $K\overline{K}$ scattering, other
channels ($2\pi,4\pi$ etc.) are opened.

For the $\pi\pi$ scattering, we obtain: $a_0^0=0.27\pm 0.06,~m_{\pi^+}^{-1}.$
Compare with results of some other works both theoretical and experimental:
the value $0.26\pm 0.05$ (L. Rosselet et al.\cite{Hyams}), obtained in the
analysis of the decay $K\to\pi\pi e\nu$ with using Roy's model;
$0.24\pm 0.09$ (A.A. Bel'kov et al.\cite{Hyams} from analysis of the
process $\pi^-p\to\pi^+\pi^-n$ with using the effective range formula;
$0.23$ (S. Ishida et al.\cite{Ishida}, modified approach to analysis of
$\pi\pi$ scattering with using Breit-Wigner forms; $0.16$ (S. Weinberg
\cite{Weinberg}, current algebra (non-linear $\sigma$-model)); $0.20$
(J. Gasser, H. Leutwyler \cite{Gasser}, the theory with the non-linear
realization of chiral symmetry); $0.26$ (M.K. Volkov \cite{Volkov}, the
theory with the linear realization of chiral symmetry).

We have here presented model-independent results: the pole positions,
coupling constants and scattering lengths. The formers can be used further
for calculating masses and widths of these states in various models.

If we suppose, that the obtained state $f_0(665)$ is the $\sigma$-meson, then
from the known relation of the $\sigma$-model between the coupling constant
of the $\sigma$ with the $\pi\pi$-system and masses
$g_{\sigma\pi\pi}=(m_\sigma^2-m_\pi^2)/\sqrt{2}f_{\pi^0}$
(here $f_{\pi^0}$ is the constant of the weak decay of the $\pi^0$:
$f_{\pi^0}=93.1$ MeV), we obtain ~$m_\sigma\approx 342$ MeV. That small value
of the $\sigma$-mass can be a result of the mixing with the $f_0(980)$ state
\cite{Volk-Yud}.

{\bf 4.} In the present model-independent approach, a satisfactory
simultaneous description of the isoscalar $s$-wave channel of the processes
$\pi\pi\to \pi\pi,K\overline{K}$ from the thresholds to the energy values,
where the 2-channel unitarity is valid, is obtained.
A parameterless description of the $\pi\pi$
background is first given by allowance for the left-hand branch-point in the
proper uniformizing variable.
Thus, a model-independent confirmation of the state, already discovered in
other works \cite{Svec}-\cite{Kaminski} (or pretending to this discovery) and
denoted in the PDG issues by $f_0(400-1200)$ \cite{PDG-98}, is obtained.
Three states ($f_0(665) - \sigma$-meson,
$f_0 (980)$ and $f_0(1500)$) are sufficient to describe the analyzed data.

The discovery of the $f_0(665)$ state solves one important mystery of the
scalar-meson family that is related to the Higgs boson of the hadronic sector.
This is a result of principle, because the schemes of the nonlinear
realization of the chiral symmetry have been considered which do without the
Higgs mesons. An existence of the low-lying state with the properties of the
$\sigma$-meson and the obtained value of the $\pi\pi$-scattering length seem
to suggest the linear realization of chiral symmetry. Remember that this
simple and beautiful mechanism works also in other fields of physics,
{\it e.g.}, in superconductivity and that the effective Lagrangians
obtained on the basis of this mechanism (the Nambu -- Jona-Lasinio and other
models) describe perfectly the ground states and related phenomena. The only
weak link of this approach was the absence of the $\sigma$-meson below 1 GeV.

Let us notice that the character of the $f_0(665)$ pole-cluster (namely,
a considerable shift of the pole on sheet III towards the imaginary axis) can
point to the unconsidered channel with which this state is, possibly, coupled
strongly, and the threshold of which is situated below 600 MeV. In this energy
region, only one channel is opened: this is the $4\pi$ channel. It is
interesting to verify this assumption, because it concerns such an important
state.

A minimum scenario of the simultaneous description of the processes
$\pi\pi\to\pi\pi,K\overline{K}$ does not require the ${f_0}(1370)$ resonance;
therefore, if this meson exists, it must be weakly coupled with the $\pi\pi$
channel, {\it e.g.} be the $s{\bar s}$ state (as to that assignment of the
${f_0}(1370)$ resonance, we agree with the work \cite{Shakin1}).

The $f_0 (1500)$ state is represented by the pole cluster which corresponds to
a flavour singlet, {\it e.g.} the glueball.

We emphasize that the obtained results are model-independent, since they are
based on the first principles and on the mathematical fact that a local
behaviour of analytic functions, determined on the Riemann surface, is
governed by the nearest singularities on all sheets.

Finally, note that in the model-independent approach, there are many adjusted
parameters (although, {\it e.g.} for the $\pi\pi$ scattering, they all are
positions of poles describing resonances). The number of these parameters can
be diminished by some dynamic assumptions, but this is another approach and of
other value.

\section*{Acknowledgments}
The authors are grateful to S.~Dubni{\'{c}}ka, S.B.~Gerasimov,
V.A.~Meshcheryakov, V.N.~Pervushin, M.K.~Volkov and V.L.~Yudichev for useful
discussions and interest in this work.

This work has been supported by the Grant Program of Plenipotentiary of Slovak
Republic at JINR.  Yu.S. and M.N. were supported in part by the Slovak
Scientific Grant Agency, Grant VEGA No. 2/7175/20; and D.K., by Grant VEGA
No. 2/5085/99.

\underline{Figure Captions}\\

\noindent
Fig.1: Uniformization plane for the  \protect$\pi\pi$-scattering amplitude.\\
\noindent
Fig.2: The energy dependence of the \protect($|S_{12}|$) obtained on the
basis of a simultaneous analysis of the experimental data on the coupled
processes \protect$\pi\pi\to \pi\pi,K\overline{K}$ in the channel with
\protect$I^GJ^{PC}=0^+0^{++}$. The data on the process
\protect$\pi\pi\to K\overline{K}$ are taken from Ref.\cite{Wickl}.\\
\noindent
Fig.3: The same as in Fig.2 but for the phase shift \protect($\delta_{12}$).

\newpage

\begin{table}[htb]
\begin{center}
Table 1: Pole clusters for obtained resonances
\end{center}
\begin{center}
\begin{tabular}{|c|rl|rl|rl|rl|rl|rl|}
\hline
{} & \multicolumn{2}{c|}{$f_0 (665)$} & \multicolumn{2}{c|}{$f_0(980)$}
& \multicolumn{2}{c|}{$f_0(1500)$} \\
\cline{2-7}
Sheet & \multicolumn{1}{c}{E, MeV} & \multicolumn{1}{c|}{$\Gamma$, MeV}
& \multicolumn{1}{c}{E, MeV} & \multicolumn{1}{c|}{$\Gamma$, MeV}
& \multicolumn{1}{c}{E, MeV} & \multicolumn{1}{c|}{$\Gamma$, MeV}\\
\hline
II & 610$\pm$14 & 620$\pm$26 & 988$\pm$5 & 27$\pm$8 & 1530$\pm$25
& 390$\pm$30 \\
\hline
III & 720$\pm$15 & 55$\pm$9 & 984$\pm$16 & 210$\pm$22 & 1430$\pm$35
& 200$\pm$30 \\
{} & {} & {} & {} & {} & 1510$\pm$22~ & 400$\pm$34 \\
\hline
IV & {} & {} & {} & {} & 1410$\pm$24 & 210$\pm$38  \\
\hline \end{tabular}
\end{center}
\end{table}

\begin{figure}
\vskip 3.cm
\centering{
\epsfig{file=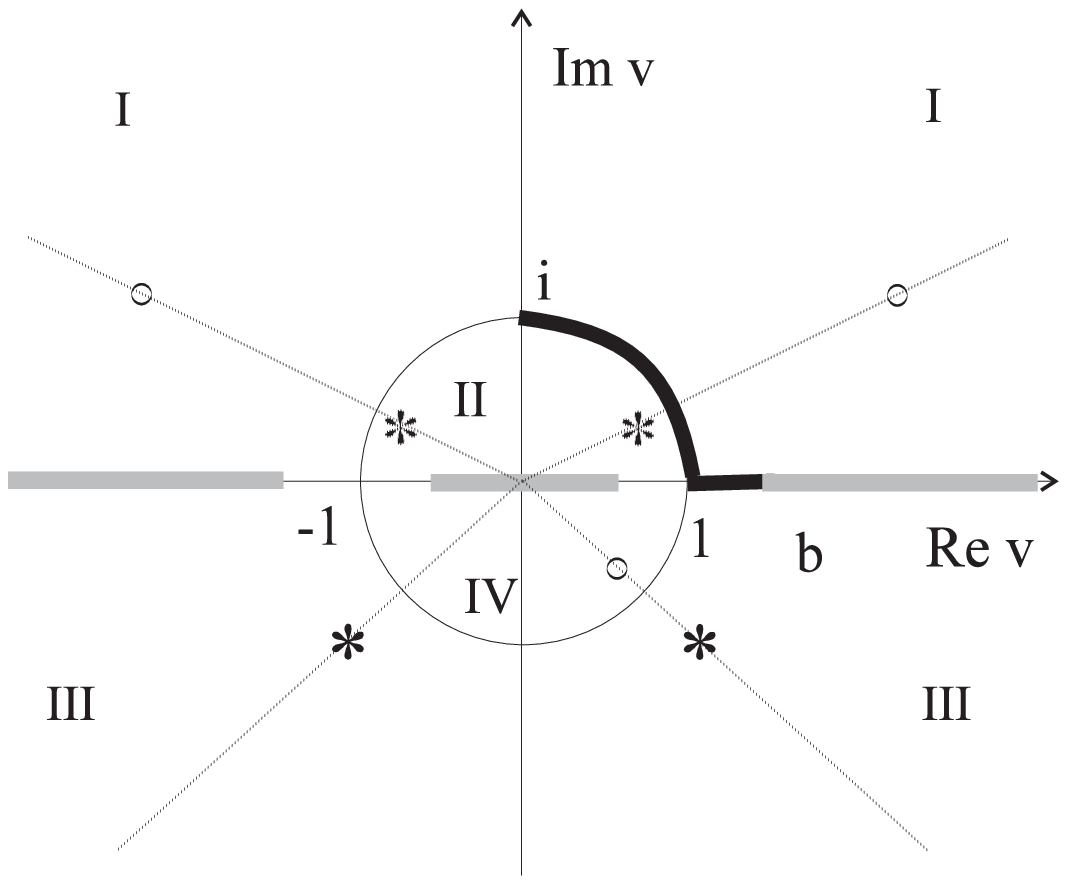, width=12cm}}
\vskip -.1cm
\caption{}
\end{figure}

\begin{figure}
\vskip 3.cm
\centering{
\epsfig{file=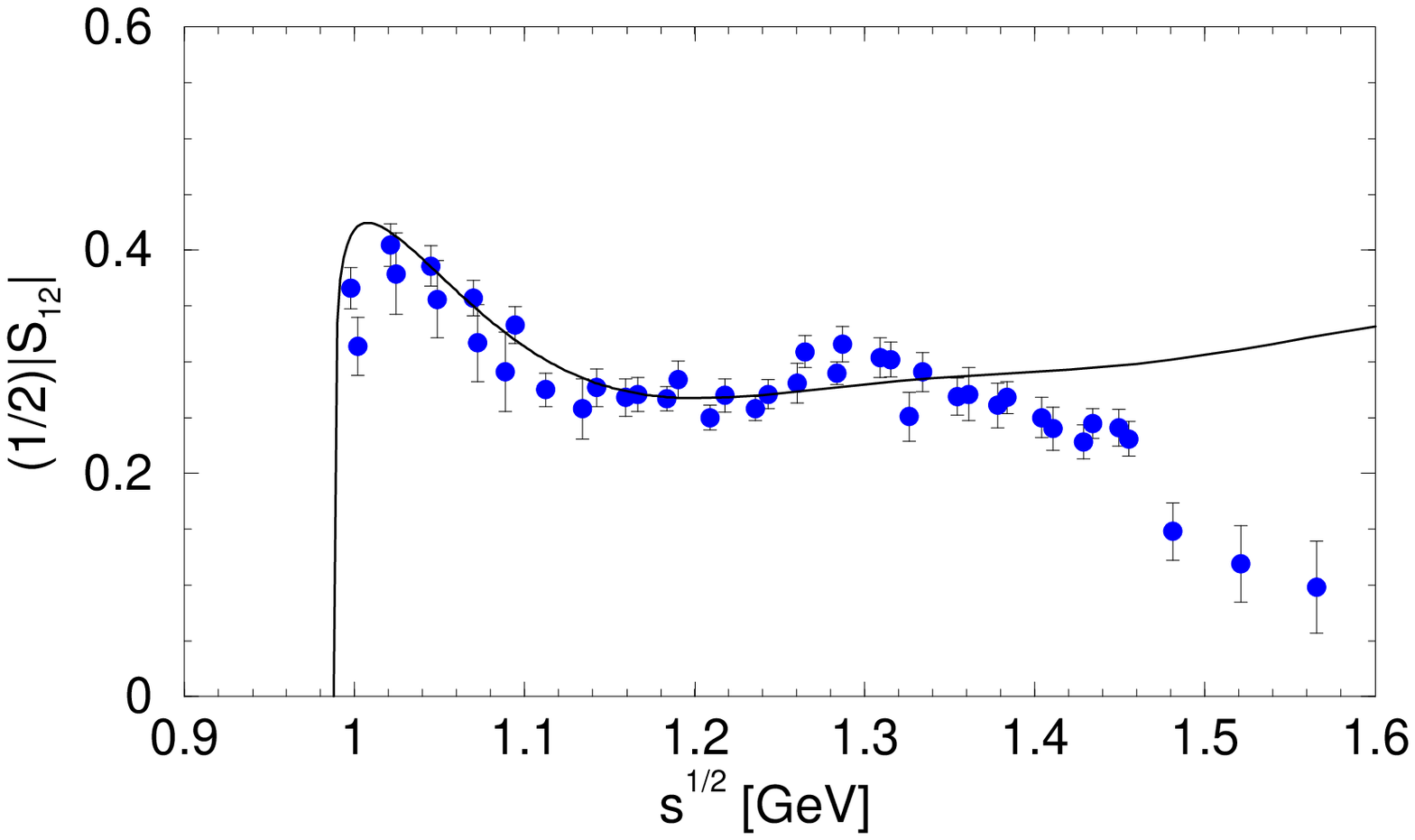, width=14cm}
}
\vskip -.1cm
\caption{}
\end{figure}

\begin{figure}
\vskip 3.cm
\centering{\epsfig{file=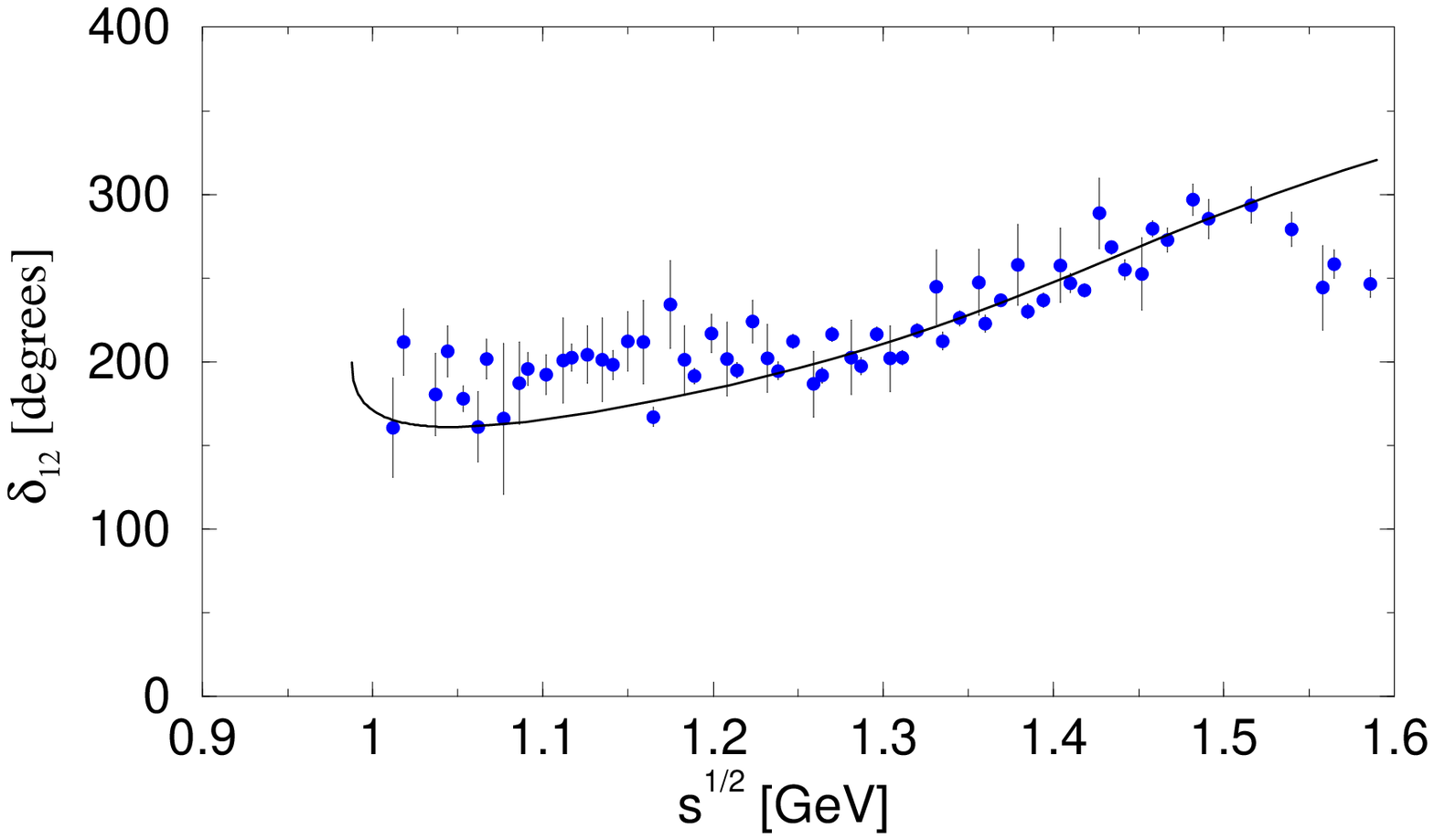, width=14cm}}
\vskip -.1cm
\caption{}
\end{figure}

\end{document}